# Would Contact with Extraterrestrials Benefit or Harm Humanity? A Scenario Analysis

Seth D. Baum,[1] Jacob D. Haqq-Misra,[2] & Shawn D. Domagal-Goldman[3]

1. Department of Geography, Pennsylvania State University. E-mail: sbaum@psu.edu
2. Department of Meteorology, Pennsylvania State University
3. NASA Planetary Science Division

**Abstract**

While humanity has not yet observed any extraterrestrial intelligence (ETI), contact with ETI remains possible. Contact could occur through a broad range of scenarios that have varying consequences for humanity. However, many discussions of this question assume that contact will follow a particular scenario that derives from the hopes and fears of the author. In this paper, we analyze a broad range of contact scenarios in terms of whether contact with ETI would benefit or harm humanity. This type of broad analysis can help us prepare for actual contact with ETI even if the details of contact do not fully resemble any specific scenario.

## 1. Introduction

Humanity has not yet encountered or even detected any form of extraterrestrial intelligence (ETI), but our efforts to search for ETI (SETI) and to send messages to ETI (METI) remain in early stages. At this time we cannot rule out the possibility that one or more ETI exist in the Milky Way, nor can we dismiss the possibility that we may detect, communicate, or in other ways have contact with them in the future.[1] Contact with ETI would be one of the most important events in the history of humanity, so the possibility of contact merits our ongoing attention, even if we believe the probability of contact to be low.

A central concern regarding possible contact with ETI is whether the contact would be beneficial, neutral, or harmful to humanity. This concern will help us decide, among other things, whether or not we should intentionally message ETI and what we should say if we do. The short answer is that we do not know how contact would proceed because we have no knowledge of ETI in the galaxy. Indeed, we cannot know for sure until after contact with ETI actually occurs. Nevertheless, we do have some information that can help us at least make educated guesses about the nature of contact with ETI. Developing and analyzing this information may help prepare us for contact and increase the probability of an outcome that we consider favorable.

---

[1] Throughout this paper we define the term "contact" broadly to include any way in which ETI has some impact on humanity. This includes human-ETI interactions that only involve remote detection or communication without any physical contact.

There have been many previous analyses of and commentaries on how contact with ETI would proceed. Unfortunately, this previous work tends to be quite narrow in the sense of only considering one or a small number of possible contact outcomes. There appears to be a tendency to jump to conclusions on a matter which remains highly uncertain and for which a broad range of outcomes are within the realm of possibility. Such narrow and hasty thought ill prepares us for actual contact. Instead, given the extremely broad range of possible contact outcomes, we would be much better prepared by identifying and thinking through a broad range of possible contact outcomes.

This paper presents a broad synthesis of available information regarding the possible outcomes of contact with ETI. Our work is in the form of a scenario analysis: we analyze many possible ETI contact scenarios in terms of whether and how they would harm or benefit humanity. In the process, we draw upon numerous prior discussions of contact with ETI that cover a broad range of possible outcomes, but tend to do so narrowly. Although contact with ETI has been discussed in the scientific literature for over fifty years [1] and in science fiction at least since *The War of the Worlds* by H. G. Wells in 1898, there has been relatively little effort to cumulatively analyze the possible outcomes compared to the synthesis presented here. To the best of our knowledge, the only previous broad synthesis is in the excellent work of Michaud [2]. The present paper has some similarities to Michaud's work but also includes several new scenarios, a different organizational structure that explicitly organizes scenarios in terms of harms and benefits to humanity, and new discussion of scenario analysis as a contribution to our understandings of and recommendations for possible ETI encounters.

Scenario analysis of ETI contact serves several purposes. First, contact scenario analysis is of strong intellectual interest to the SETI and METI community and others, given the nuances and challenges involved in imagining an ETI we have never observed. But this scenario analysis is of practical value as well. An individual scenario is a narrative of a possible outcome of, in this case, contact between humanity and ETI. Such scenarios can help us train our minds to recognize patterns in actual outcomes. By "training our minds" we mean simply that our minds grow accustomed to thinking about, identifying, and analyzing specific scenarios and variations of them. The training process is thus simply reading and reflecting on the scenarios and the encounter patterns found in them. The patterns of an actual encounter may resemble the analyzed scenarios even if the specifics differ from the scenario details. By training our minds in this way, we build our capacity to analyze and respond to actual contact with ETI. The scenario analysis presented here thus holds practical value in addition to the noteworthy intellectual insights that come from considering how contact with ETI might proceed. Additionally, by considering a broad range of possible contact scenarios, including some that might seem unlikely, we improve both the range of patterns our minds are trained for and the breadth of intellectual insight obtained. This sort of broad scenario analysis can thus be an especially fruitful process.

We organize ETI contact scenarios into three basic categories based on whether the consequences would be beneficial, neutral, or harmful to us. Although the possibilities surely fall along a spectrum along these lines, we believe these three bins represent a useful categorization scheme. As defined here, beneficial contact would be desirable for humanity; neutral contact would cause indifference for humanity; and harmful contact would be undesirable for humanity. A relatively large number of the scenarios we consider fall within the harmful-to-humanity category. We thus further divide these scenarios into two sections in which ETI are either intentionally or unintentionally harmful. Note that the large number of harmful-to-humanity scenarios does not imply that contact with ETI is likely to harm humanity. Quantitative estimates of the probabilities of specific scenarios or categories of scenarios are beyond the scope of this paper. Here we focus instead on the breadth and form of the possible modes of contact with ETI. Before developing these scenarios, we present some background information of relevance to the discussion that follows.

## 2. Relevant background

Some background information is relevant to many of the ETI contact scenarios discussed in the rest of the paper and is thus worth considering separately and in advance of the scenarios. This background concerns why we have not yet detected ETI (i.e. the Fermi paradox), the challenge of interstellar communication, why ETI are likely to be more technologically advanced than humanity, what we can learn about the ethics held by ETI from the study of ethics held by humans, and the possibility of heterogeneity within an ETI population.

### 2.1 The Fermi paradox

So far, no extraterrestrial civilization has been unequivocally observed by humans. Nearly 50 years of listening for ETI transmissions has found no artificial signals in space [3-4], and the search for ETI artifacts in the Solar System has also produced null results [5-7]. However, a simple back-of-the-envelope calculation initially performed by physicist Enrico Fermi suggests that ETI should be widespread throughout the galaxy [8]. Indeed, an advanced ETI civilization could easily colonize the galaxy to form a Galactic Club among intelligent societies, a concept popular in science fiction (such as the "United Federation of Planets" of *Star Trek* fame) that in the nonfiction literature dates back at least to Ronald Bracewell [9]. This conspicuous absence of extraterrestrials is often referred to as the Fermi paradox [8] or the Great Silence [10] and raises the question: if ETI should be widespread, then *where are they?* A number of resolutions to the Fermi paradox have been proposed and explored [11-12], and three paradox resolutions are worthy of consideration in our discussion.

One resolution to the Fermi paradox is that life, or at least intelligence, is rare and thus sparsely distributed throughout the galaxy. This rarity could be because few intelligent civilizations form [13] or because intelligent civilizations tend to have short lifetimes, perhaps because they quickly destroy themselves [14-15]. If intelligence is rare, then it is quite unlikely that humanity would have detected ETI. In the extreme case, humanity is the only intelligent civilization in the galaxy or even in the universe. Along the same lines, other intelligent civilizations may be beyond the physical limits of contact even if they do exist [15-17]. These scenarios are of limited value to this paper because they imply that contact with ETI is impossible.

A second possible resolution to the Fermi paradox derives from the challenges of expanding rapidly throughout the galaxy. Perhaps rapid expansion is unsustainable at the galactic scale, just as rapid expansion is often unsustainable here on Earth. This suggests that the absence of extraterrestrials might be explained by the fact that exponential growth is an unsustainable development pattern for intelligent civilizations [18], a response to the Fermi paradox known as the Sustainability Solution [19]. According to the Sustainability Solution, rapidly expanding civilizations may face ecological collapse after colonizing the galaxy, analogous to the fate of Easter Island [20]. On the other hand, the galaxy could be teeming with ETI that expand too slowly to have reached Earth yet [21]. These slowly expanding ETI civilizations could still be detected by us or send us messages, and their nature as slow expanders has some implications for contact scenarios.

A third response to the Fermi paradox suggests that ETI are actually already widespread throughout the galaxy but are somehow invisible to us. The ETI could be unintentionally invisible, if it just happens to take some form that is undetectable to or otherwise undetected by humans. Alternatively, the ETI could be intentionally invisible. The intentional form of this solution is sometimes known as the Zoo Hypothesis [22] because it implies that ETI are treating Earth like a wildlife preserve to be observed but not fully incorporated into the Galactic Club. This idea has been popularized through the *Star Trek* series as the "prime directive" for non-interference with a primitive culture. The Zoo Hypothesis thus implies that ETI could make contact with humans at any time. Perhaps such stealthy ETI will reveal themselves once Earth civilization has reached certain milestones. They may be waiting until we have reached a sufficient level of sophistication as a society such as the start of a METI program or the discovery of light speed travel [22-23], or they could be applying a societal benchmark such as sustainable development or international unity. The possibility that the Zoo Hypothesis explains the Fermi paradox has several important implications for contact scenarios.

**2.2 Interstellar communication**

Even if ETI exist in the nearby galactic vicinity, this does not necessarily imply that communication with them will be possible or straightforward. One major challenge is selecting the frequency at which to broadcast and listen [24]. The electromagnetic spectrum consists of a continuum of wavelengths for communication that includes radio, microwave, infrared, visible, ultraviolet, and x-ray bands. Searching this entire range is a monumental and nearly impossible task, so we choose particular wavelengths that seem more probable for interstellar communication. For example, the 21 cm hyperfine transition of neutral hydrogen was the first suggestion for a communication wavelength [1]. The *water hole* at a wavelength of 18 cm is another popular choice for SETI [24], and recent analysis has suggested that we shift our focus toward higher frequencies [25]. However, because there is an infinite number of wavelengths for interstellar communication, we must acknowledge the possibility that ETI may be transmitting or listening at wavelength ranges that we have not yet considered. The possibility also remains that ETI do not use electromagnetic radiation for communication but instead have discovered some other method (possibly something more efficient or effective) for exchanging information across astronomical distances.

Communication via electromagnetic radiation is limited by the time required for a signal to reach its destination, i.e., the speed of light. On Earth, electromagnetic communication is nearly instantaneous because of the short distances involved. However, galactic communication occurs over astronomical distances so that even a message traveling at light speed will take a long time to reach its destination. For example, communication with ETI on a planet just 50 light years away–which is relatively close by galactic standards–will still take place on a timescale of 100 years. As Sagan [15] notes, this makes communication with ETI an intergenerational project: effective communication across astronomical distances will require unprecedented cooperation that spans several human lifetimes. This difficulty in communicating across such vast distances also might limit the ability for ETI to engage in interstellar warfare for the simple reason that the communications problem renders such warfare too logistically difficult to coordinate [26]; peaceful endeavors such as the formation of a Galactic Club may face similar logistical challenges. Such physical limits on interstellar communication by ETI are in turn limits as to how ETI could more generally come into contact with and affect humanity.

Another implication of these long communication times across the galaxy is that ETI might become alerted to our presence without us realizing it. Communication with electromagnetic waves on Earth has been used for nearly one hundred years, during which time our radio shows, television programs, and mobile phone conversations have isotropically leaked into space. If ETI search for us just as we search for them, i.e. by scanning the sky at radio and optical wavelengths for any type of interstellar communication [4], then they might detect our *leakage* signals. Advanced ETI within 100 light years could receive our earliest radio transmissions; those less than 50 light years away could watch our television shows [27]; and those less than 10 light years away could receive our earliest intentional METI attempts [28]. Thus, the radiation that has been unintentionally leaking and intentionally transmitted from Earth may have already alerted any nearby ETI to our presence and may eventually alert more distant ETI. Once ETI become alerted to our presence, it will take at least as many years for us to realize that they know we are here. During the intervening time, ETI can respond to our presence or prepare for contact in ways that we would have no knowledge of or influence on.

Even if humanity can successfully exchange signals with ETI, there is no guarantee that the information will be successfully communicated. In order for information to be exchanged, it is also necessary that humans and ETI understand the contents of each others' messages. It will likely be difficult at first to communicate anything subjective about human experience, emotions, and expressions, so mathematical conversation may comprise our first few exchanges with ETI [29]. It may eventually be prudent to develop a framework for METI so as to increase the probability of successful communication anytime a transmission is sent from Earth [30]. Perhaps such schemes will succeed in effectively communicating with ETI. However, our extreme ignorance about the nature of any ETI means that we cannot rule out the possibility that we will fail or at least severely struggle to exchange information with them.

**2.3 The advanced nature of extraterrestrials**

If contact between humans and ETI is possible, then it is important to consider the capability of ETI to cause us benefit or harm. This information is important across nearly the full breadth of contact scenarios. Although we cannot know the level of technological sophistication achieved by ETI, we do have a compelling reason to believe that ETI would be significantly stronger than us and therefore highly capable of causing our total destruction. This point has been raised repeatedly throughout the literature [1,4,14-16,31-33].

The reason to believe that ETI would be more advanced is because humans and human technology are relatively recent phenomena in the history of Earth. We have only had radio communication for about a century, or just a few generations, which suggests that advanced technology can develop quickly compared to evolutionary timescales. Following this reasoning, it is likely that any extant ETI has been around much longer than us and would have developed far greater technological abilities than we could imagine for ourselves. Even if an ETI is younger than us, the very ability to contact us would likely imply progress beyond that which our society has obtained. We have not yet figured out how to achieve interstellar communication or travel; a society that has these capabilities is almost certainly more technologically advanced than we are. If their communications are directed toward a general audience and not only intended for humans or Earth, then they may also be more advanced in their ability to communicate across cultural barriers. This is reminiscent of Arthur C. Clarke's insight that "any sufficiently advanced technology is indistinguishable from magic". If ETI are indeed more advanced, then any form of contact will likely proceed according to the ETI's desires, whatever those might be [34]. For example, we are almost guaranteed to lose in a fight between us and them, and there is a strong likelihood that such a loss would be so severe that we would cease to survive as a civilization. On the other hand, if ETI decide to use their superior abilities to help us, then they may be able to help solve many of our problems.

**2.4 Extraterrestrial ethics: Selfishness and universalism**

As noted above, if ETI are significantly more advanced than humanity, then the outcome of contact may depend primarily on ETI desires. However, this leaves open speculation as to the specific desires of ETI and raises the question of what ethical framework they follow. Much can be said about ETI ethics. Here we focus on one key aspect: selfishness vs. universalism. In rough terms, a selfish ETI is one that desires to maximize its own self-interest, whereas a universalist ETI is one that desires to maximize the interests of everyone, regardless of which civilization they are part of. But this is a crude explanation of selfishness and universalism; more precision is needed for our purposes in this paper.

As a starting point, it is helpful to think of ETI as trying to maximize some sort of value function.[2] Specifically, they are trying to maximize intrinsic value, which is something that is valuable for its own sake. Intrinsic value contrasts with extrinsic value, in particular instrumental value, which is valuable because it causes additional value. One can place intrinsic value on many different things, such as life, ecosystems, happiness, knowledge, or beauty. Human ethics is often anthropocentric in the sense that it places intrinsic value only on human phenomena, such as human life, human happiness, or other human factors. Such anthropocentrism is selfish on a civilizational scale because it involves humans only placing intrinsic value on the interests of their own civilization. In contrast, a universalist ethical framework would place equal intrinsic value on certain phenomena regardless of which civilizations possessed these phenomena. For example, a universalist civilization that places intrinsic value on life will place equal intrinsic value on all life, regardless of which civilization (or non-civilization) the life is part of. In this case, the civilization will try to maximize the total amount of life, regardless of whose life it is maximizing. If instead it places intrinsic value on some phenomenon other than life, then it will try to maximize that phenomenon wherever it occurs.

---

[2] The discussion here is derived from the more detailed discussion found in the work of Baum [34].

Conflicts between humans are often, though not necessarily always, rooted in selfishness. These conflicts include struggles for power, land, resources, prestige, and many other instruments of self-interest. Even when human conflicts have overtones of being for some higher purpose, such as for liberty or against oppression, the basic desire for the survival and flourishing of the self often remains a core motivation. Likewise other conflicts we see throughout the sentient animal kingdom appear to be motivated by the desire for instruments of self-interest such as survival, food, or territory [35]. While non-sentient species (animal or otherwise) may also appear to act in their own self-interest, it is inappropriate to attribute intent to them because intent is presumably a property of sentience.

It is worth noting that the analysis in this paper is in a sense selfish in that it focuses on benefits and harms *to humanity*. Throughout the paper, we do not consider how contact with humanity could benefit or harm either the ETI or any other entities affected, including other entities on Earth and elsewhere in the galaxy. By focusing on benefits and harms to humanity, we do not intend to advocate for a selfish ethics. Instead, this focus is simply an expository tactic aimed at keeping this article reasonably concise. In our view, consideration of impacts of contact to nonhumans is important and would be well worth considering in future work.

**2.5 Possible ETI heterogeneity**

The scenario analysis presented throughout this paper assumes that any given encounter will follow one general trajectory. The encounter might benefit, be neutral to, or harm humanity for a certain reason, but the encounter would only have one of these outcomes and follow one general trajectory to reach this outcome. This follows from the idea of a homogenous ETI, i.e. an ETI with one defining attribute or combination of attributes that dominates the encounter. The attribute could be the ETI's strength, ethics, politics, or something else. If it is the case that the ETI has one defining attribute or combination of attributes, then it is reasonable to expect one general trajectory for the encounter. However, this requires a homogenous ETI population.

It is possible that an ETI would have a heterogeneous population instead of a homogenous one. Evidence for this can be found in the human population, which features a highly diverse mix of technological abilities, ethical views, national identities, and other attributes. For example, in the event of an ETI encounter, humanity may be fiercely divided on whether to respond peacefully or with protective aggression. ETI may be similarly divided. At a minimum, humanity's diversity provides proof of the principle that intelligent civilizations can be heterogeneous.

The possibility of ETI heterogeneity suggests that an encounter might not follow one general trajectory but instead could have multiple trajectories in series or perhaps even in parallel. For example, an encounter could rapidly change form if a shift in power occurred within the ETI leadership. Or, we might receive mixed signals from the ETI if it lacks a single unified leadership structure; perhaps several ETI factions or nations that originate from the same home world will make contact with us, each in pursuit of different objectives. The possibilities of ETI heterogeneity and multiple trajectories are worth keeping in mind when considering the specific encounter scenarios that could occur.

Having considered these points of background information, we can now proceed to specific scenarios of contact between humanity and ETI. An overview of these scenarios is provided in Fig. 1.

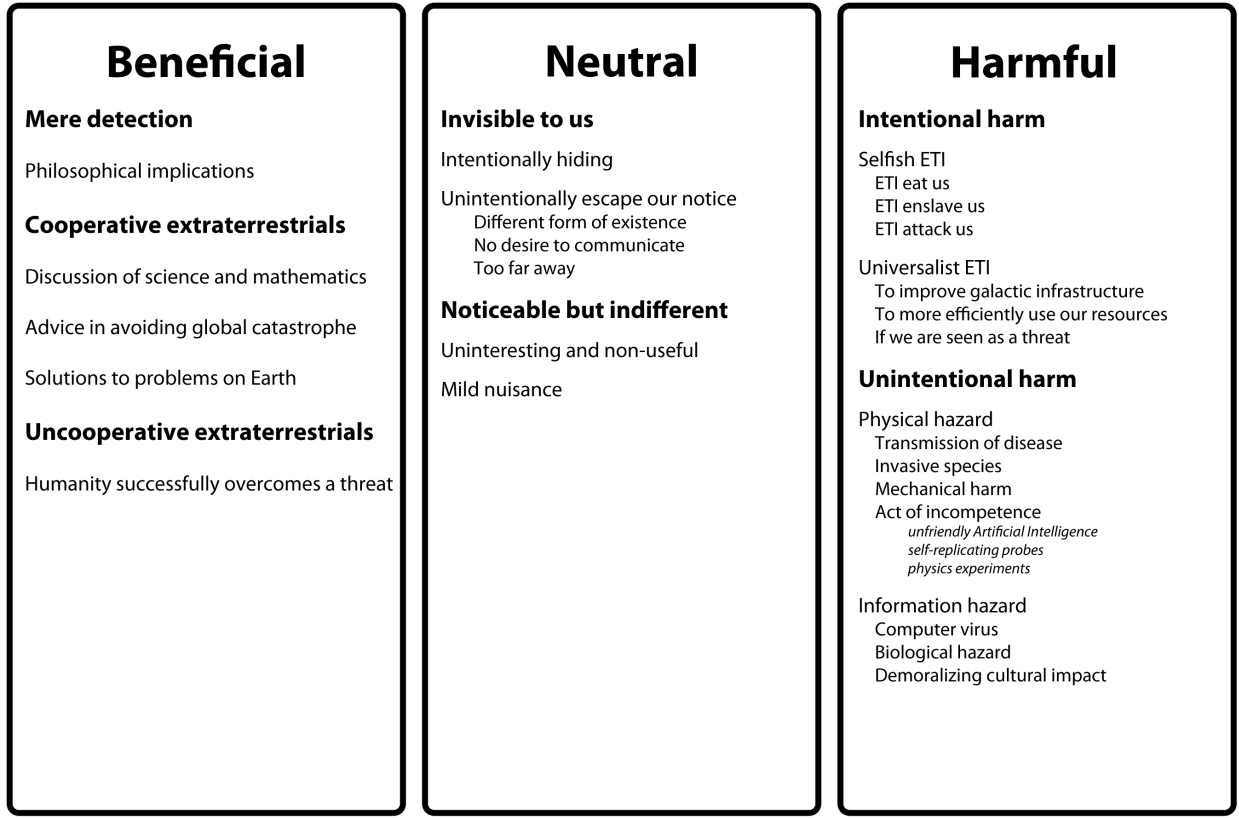

**Fig. 1.** Overview of the contact scenarios presented in this paper.

### 3. Beneficial to us

The most optimistic scenarios assume that contact with ETI would somehow benefit humanity (Figure 1, left column). These scenarios are broadly popular: survey results have shown that many people across the world anticipate that contact with ETI will benefit humanity in some way [36-37; see also 38]. The nature of this benefit could range from simple remote detection of intelligent life elsewhere to more extensive contact with cooperative ETI. There is also at least one set of scenarios in which we benefit from contact with uncooperative ETI. While we cannot know whether an ETI would be cooperative, we present some reasons to suspect that they would be cooperative by developing in some length an argument based on the Sustainability Solution to the Fermi paradox.

### 3.1 Mere detection

Mere detection of ETI refers to scenarios in which the entirety of contact is limited to the discovery that ETI exist. In other words, we detect the presence of ETI and thus can confirm their existence but have no further contact. This means no communication, direct contact, or any other possible contact mode. Here we argue that mere detection would provide a nontrivial benefit to humanity.

If ETI do exist within the galaxy, then confirmation of their presence would have profound implications for human science, philosophy, religion, and society. This point has been noted repeatedly throughout the literature [15,33,39-41]. Indeed, ongoing SETI activities are based to a large degree on the premise that humanity wants to learn about ETI. One reason for this is that the discovery of ETI would answer the deep and longstanding philosophical question of whether we are alone in the universe. This in turn relates to the question of our role in the universe as intelligent beings. Humanity has a strong interest in obtaining answers to these major questions and thus would benefit tremendously from the mere detection of ETI.

Some people might consider mere detection to be harmful to humanity. These people include those with religious perspectives and other worldviews that depend on the idea of humanity (and Earth-life more generally) playing a unique and privileged role in the universe (e.g., [42-43]). The detection of ETI could challenge these worldviews and therefore be perceived as harmful by those who hold such beliefs. However, this perception of harm depends on a philosophical mistake. The existence of ETI in the universe is independent of whether or not they have been detected by humanity. It is the existence of ETI that challenges such worldviews and not the act of detection. If ETI do in fact exist, then the harm has already been done in the sense that such worldviews are already invalid. Detection simply alerts us to this invalidity. This alert itself might be classified as a benefit or harm, because of its affects on the wellbeing of those whose worldviews are challenged with the discovery of ETI, but this is seemingly a lesser matter than the broader benefits of mere detection.

More troubling is the possibility that detection could initiate or exacerbate conflicts in our society. The conflict could be over how to interpret or reply to such a discovery. There are already disagreements over how to message to ETI, whether or not we should, and who should speak for humanity; such disagreements would become much fiercer if ETI were detected. Meanwhile, the groups whose worldviews would be challenged could respond in harmful ways if they feel threatened, nullified, or otherwise worsened by the discovery or the intent to reply. While we hope that detection would unify humanity towards positive outcomes, the opposite result remains entirely possible.

While mere detection of ETI would be beneficial for the insight it offers, these benefits could be limited. That is, mere detection would leave much of humanity's situation intact. Perhaps mere detection would be on par with the Copernican revolution in that it would change human thought but not radically alter our geopolitics [44]. So while mere detection may offer net benefits, these benefits are likely not very large, especially relative to the benefits and harms found in many other contact scenarios.

Regardless of their magnitude, the impacts of mere detection serve as a baseline set of impacts for almost all other contact scenarios. This is because nearly all other contact scenarios involve detection along with other forms of contact. The exceptions here are contact scenarios that do not involve detection, which include scenarios in which ETI manipulate our world (in good ways or bad) while hiding and scenarios in which ETI destroy us without our having the opportunity to notice the ETI. These scenarios are discussed further below.

Even if we receive no more than a simple greeting or passive artifact from a distant ETI civilization, it will at least tell us that life has developed more than once in the galaxy and that human-like technology to broadcast across space has been invented elsewhere. Advanced ETI may have little to no interest in a society as primitive as Earth, but if they do acknowledge our presence and initiate communication, then even this knowledge will benefit humanity.

### 3.2 Cooperative extraterrestrials

If contact with ETI involves more than mere detection, then it is possible for humanity to receive additional benefits by cooperating with the ETI. The nature of these benefits depends on the degree of ETI cooperation – that is, it is unlikely that uncooperative ETI would benefit humanity. This is because ETI are likely to be much more advanced than humanity and would therefore be capable of dictating the terms of contact. Thus cooperative ETI would have the ability to bring benefits to humanity, just as uncooperative ETI would likely harm humanity.

An initial scenario of cooperative ETI involves friendly and informative communication between our respective civilizations. Assuming ETI are sufficiently interested in humanity (which is not guaranteed, given that they would likely be much more advanced), they may choose to maintain communication at length to discuss mathematics, physics, and chemistry [29] and to learn more about Earth life. It is reasonable to assume that the general principles of physics and chemistry apply everywhere in the galaxy, even if mathematical descriptions of these physical phenomenon differ among intelligent civilizations. This type of dialog with ETI may require that we first develop a common mathematical language using physical observables that are known by both civilizations (such as properties of neutral hydrogen). In a more remarkable and unlikely case, we may learn that ETI occupy some region of space where different or unknown physical principles apply, which would certainly be a unique discovery for humanity. Thus through such a conversation we may come to acquire a deeper understanding of mathematics or science, and we may also discover specifics about the ETI home world or ETI biology. As with mere detection, such contact would have considerable intellectual benefits, though here the benefits would be larger – potentially much larger.

Depending on the nature of information shared through communication with ETI, there could also be more in the way of practical, non-intellectual benefits. An advanced ETI may be capable of solving a great many of humanity's problems, such as world hunger, poverty, or disease. Benevolent ETI may even design their first message to contain information on how to avoid technological catastrophe in order to help less developed civilizations succeed [45]. From humanity's perspective, this is the best-case scenario for ETI contact. However, while we suspect that the basic principles of physics and chemistry apply across the universe, it is somewhat less likely that ETI knowledge would be useful in addressing social issues on Earth. The usefulness of ETI knowledge, combined with the willingness of ETI to employ it on our behalf, plays an important role in the benefits that a cooperative ETI would bring to humanity.

We do not know if ETI would be cooperative, but we have several reasons to suspect that they would be. Noncooperation can be a risky and harmful strategy, and noncooperative civilizations may tend to have shorter lifetimes as their noncooperation eventually leads to their demise. For this reason, a long-lived civilization that explores the galaxy may have transcended any aggressive patterns out of the need to maintain long-term survival [36,46]. It is also possible that intelligent civilizations may inevitably develop cooperative tendencies as part of their evolutionary process [44,47]. However, there are also reasons to suspect that evolution would proceed along different, less desirable trajectories [48].

Another reason to suspect that ETI would be cooperative follows from the Sustainability Solution to the Fermi paradox. A corollary of the Sustainability Solution is that extant ETI civilizations in the galaxy may be less prone to violence and destruction in the event of contact. This corollary follows from the tendencies of sustainable human populations.

On Earth, sustainable human populations tend to be more protective of their ecosystems. This protectiveness can be for either of two reasons. First, humans can protect ecosystems for their own benefit. This protection is known as conservationism and involves humans placing intrinsic value on themselves. Second, humans can protect ecosystems for the ecosystems' benefit. This protection is known as preservationism and involves humans placing intrinsic value on the ecosystems. (See [49] for a similar approach to environmental ethics in the context of terraforming Mars.) In either case, human populations that follow a sustainable mode of development are less likely to expand for lack of resources, although they may choose to explore out of sheer curiosity. ETI populations may be similar in this regard [50]. Thus, if exponential growth is in fact unsustainable on the galactic scale as Haqq-Misra and Baum [19] suggest, then we are much more likely to encounter a long-lived ETI civilization that follows a sustainable development pattern. Such a civilization may have no need to consume Earth systems (or humans) because they will have already found a way to effectively manage their resources over long timescales. Therefore, the possible unsustainability of long-term rapid expansion decreases the probability that ETI will destroy us. However, there is a scenario in which sustainable ETI would destroy us – specifically if the ETI is expanding at the maximum rate possible given its sustainability constraints. This "maximally expansive" scenario is one of the "harmful to humanity" scenarios discussed below.

### 3.3 Uncooperative extraterrestrials

Given that ETI are likely much more advanced than human civilization, contact with uncooperative ETI seems likely be harmful to humanity. Harm from uncooperative ETI is discussed in detail in Section 5. However, there are certain scenarios in which contact with uncooperative ETI would benefit humanity. These are scenarios in which ETI attempts to harm us but fails. Perhaps the ETI, no matter how strong or powerful, just happen to be vulnerable to something humanity has. This is illustrated in the conclusion to *The War of the Worlds*, where the invading Martians are rendered helpless by infection by Earth microbes. Or perhaps humanity somehow goes against the odds and defeats the ETI. This latter scenario is widespread throughout science fiction, including in major Hollywood films such as *Independence Day* (1996). In these scenarios, humanity benefits not only from the major moral victory of having defeated a daunting rival but also from the opportunity to reverse engineer ETI technology. A final scenario involves a second ETI learning of our situation and coming to our rescue, again leaving us better off than we were to begin with. Scenarios such as these might make for quality entertainment, but they also appear rather unlikely. Still, such contact scenarios are possible and thus worth including in this analysis.

### 4. Neutral to us

Another set of scenarios involves contact with ETI that are neutral toward us (Figure 1, center column). Neutral here means that humanity is indifferent to contact with ETI: we are just as well-off with it as we are without it. There are two fundamental ways in which ETI could be neutral. The most straightforward way is that ETI have no impact on us at all. Here it is important to recognize that ETI would have a profound impact on humanity if we simply become aware of its presence – that is, through mere detection, as discussed above. Indeed, the discovery of ETI could well be the most profound and important discovery that humanity has ever made. Thus, for ETI to have zero impact on us, they must go undetected. In other words, ETI will have no impact on us only if they remain invisible to us.

The other way in which ETI could be neutral is if they have an impact on humanity, but the cumulative effect of that impact is neutral. In this case, humanity becomes aware of the presence of ETI. As discussed above, detecting ETI is generally considered beneficial for humanity. Therefore, if we detect ETI and are neutral toward them, then there will have to be some harm in order to offset the benefit of contact. It is unlikely that this harm would precisely offset the benefit of detecting ETI (and any other benefits that might come with contact), so here we consider scenarios in which the offset is of approximately the same magnitude, which results in a net impact that is roughly neutral.

**4.1 Invisible to us**

There are several scenarios in which ETI could be invisible to us in the sense that we do not detect the presence of any ETI. All of these scenarios assume that ETI do in fact exist, but we do not detect their presence, perhaps because we are physically unable to do so. As far as humanity is concerned, invisible extraterrestrials could be no different than non-existent extraterrestrials if they both have no impact on us. This scenario would be completely neutral to us. However, it is not necessarily the case that an invisible ETI would have no impact on us.

One invisibility scenario involves ETI that intentionally hide from us. This corresponds to the Zoo Hypothesis of the Fermi paradox. ETI could have the capability of hiding from us given the likelihood of their superior technology, and there are many ways that ETI could remain undetected by us if it chooses to do so. The simplest approach would be to hide among the asteroids and observe us at a distance [51-54]. In this case, such ETI will cease to be invisible to us when we have searched enough of the asteroid belt to detect signs of their presence, such as mining on asteroids [55-57], excess infrared radiation from spacecraft [7,58], or intelligent conversational space probes [59]. A more sophisticated approach would eliminate all outgoing electromagnetic signals by to hide any signatures of its presence, and ETI with even greater technological prowess could engineer a virtual planetarium surrounding Earth so that we are forced to observe an empty universe [60].

Depending on the form of the intentional hiding, the scenario need not be strictly neutral. Deardorff [23] argues that hidden ETI may actually be beneficial because they know we are here and presumably check up on us from time to time. Perhaps they do have our best interests in mind and will initiate friendly contact when we begin a long-term METI program [23] or when we start conversing with an intelligent space probe [59]. This scenario may even have some implications for human behavior that are somewhat parallel to scenarios in which humanity is actually the manifestation of a computer simulation [61]. A sustainable ETI may be hiding from us to see if we can turn into a sustainable society on our own before we gain the ability to travel between stars. Such a society would be temporarily neutral to us, but potentially harmful or beneficial to us in the long term.

Another possibility is that ETI would unintentionally escape our notice [32]. Even if they took no extraordinary measures to remain concealed, ETI that pass by Earth may draw as much attention from humans as a passing-by scuba diver would alert a sea anemone by taking a photograph. This could be because ETI take a different physical form than Earth life – a form that we are unable to recognize – or because their technology is unobtrusive enough that we fail to take notice. Although it is common to assume that extraterrestrial life will most likely be carbon-based and require liquid water, there are a number of suggestions for more exotic configurations of life. These include alternative biochemistries based on alcohol solvents or silicon [62-63], a shadow biosphere that invisibly coexists with the life we know [33,64], pure energy beings that lack a physical form, and even residence between multiple universes [65-66]. In the same way, we may fail to notice ETI messages that are transmitted at a different wavelength range than we typically listen to. In such a scenario, ETI are actively attempting to communicate with us, but we miss the message because our search efforts are less than comprehensive. ETI may be interested in observing the Earth system for scientific purposes or may simply be galactic tourists passing through the Solar System. But as long as they refrain from significantly interfering with humans or our environment, these ETI provide no threat or benefit to our existence.

It is also plausible that nearby ETI simply have no desire to communicate with us. Non-expansive ETI that pursue a sustainable development pattern may also find all the contentment and meaning they need on their own planet so that they have no desire for interstellar communication [41]. They may have taken up transcendental spiritual practices that focus their efforts inward rather than outward [39], or they might limit their space exploration to passive interstellar probes [31,67-70]. Perhaps ETI actually do inhabit nearby star systems and detect our radio leakage but have no plans to send a response until we send them a more intentional message [23,39]. They may be unimpressed with the quality of our broadcasts, or they may choose to conserve their resources and decide that interstellar communication is too expensive. For our purposes here, these non-communicative ETI are invisible all the same.

Finally, we must acknowledge the possibility of scenarios in which ETI are too far away for communication. It may be that ETI have no desire to maintain long-term communication with us, but they also may be too physically distant from Earth to consider communication [15-17]. An ETI broadcast from another galaxy, for example, may not have yet reached Earth and would probably be too faint to detect with modern technology. Likewise, ETI that live beyond 100 light years from Earth would not have detected our radio leakage and may not yet know of our presence. Even if intelligent life is common in the universe, it may still only arise once or twice within a galaxy [13] so that the likelihood of interstellar communication is diminished. Then again, the galaxy may be full of non-expansive ETI that may still embark upon interstellar radio communication but are too far away for us to have yet received their messages. Human expansion in space may lead to eventual contact with non-expansive ETI, but aside from this possibility, non-expansive ETI will remain invisible to us and have little influence on humanity.

**4.2 Noticeable but indifferent to us**

It is possible that humanity could succeed in identifying ETI in the galaxy, only to find that we are indifferent to the cumulative experience. This may seem unlikely, given that the discovery that extraterrestrials exist elsewhere in the galaxy would have wide-reaching implications. Nevertheless, there are several scenarios in which our evaluation of the encounter could be one of indifference.

As an initial scenario, suppose that planet-finding missions successfully identify an extrasolar terrestrial planet orbiting a Sun-like star with an atmospheric composition similar to Earth [71]. Follow-up observations with radio telescopes reveal unintentional electromagnetic leakage coming from the planet, which suggests the presence of intelligent life. However, suppose further that we decode this leakage to find no more than the ETI equivalent of old television shows and obscure military transmissions. These broadcasts may contain next to nothing in terms of information usable by humans, and the public may quickly lose interest in non-responsive ETI with uninteresting messages [36]. Even active ETI broadcasts that are targeted toward Earth may contain information that we find useless or esoteric. Remote observation of an ETI planet may also reveal strikingly different chemical compositions between their world and ours. ETI that originate from a gas giant planet, for example, may have followed a completely different evolutionary trajectory that leaves little room for biological similarity between us and them. Communication with such ETI may provide little useful information for humans. After all, an ETI society that eats only hydrogen might not have any practical information relating to development issues on Earth, and the vast difference in biology might render them unable to communicate with us at all. If the search for life finds that the galaxy is in fact teeming with ETI, then uninteresting ETI planets such as these will likely fall to low priorities for making contact.

Another scenario involves us finding that contact with ETI creates a mild nuisance or requires more effort than we would like to spend. The film *District 9* (2009) highlights a contact scenario where we discover a helpless ETI crew that requires human assistance in order to survive [72]. Placed in a temporary refugee camp, the ETI in *District 9* display a wide range of temperaments, but their overall presence annoys the humans because the ETI refugee camp seems to be a permanent fixture. Similarly, we may discover through remote messaging that ETI need our help but provide little in return, so that contact with ETI eventually begins to drain human resources. Under different circumstances, perhaps ETI make contact with Earth to welcome us into the Galactic Club but only after we complete a set of required bureaucratic tasks. ETI that make contact with Earth will certainly be more technologically advanced than humans today, so it is entirely plausible that the requirements to enter a Galactic Club will be beyond our abilities. In these scenarios, contact with ETI benefits humanity by confirming the presence of life elsewhere, but the consequences of contact are sufficiently disruptive, annoying, or complicated that human civilization remains indifferent.

A final scenario involves disagreement within human civilization regarding whether or not contact has occurred. The simplest conditions for this to occur would be if we received a message from ETI that cannot be unambiguously decoded. No SETI signal has yet been identified as extraterrestrial in origin, and if we do ever stumble upon an actual ETI broadcast then there could be a long and tedious process to demonstrate its authenticity. Less probable modes for this form of contact have been explored in films such as *Contact* (1997) and *K-PAX* (2001) in which the nature of the ETI is only realized by a handful of humans and dismissed by the rest. If our detection of ETI lacks an obvious and unambiguous signal, then different humans – even including different ETI researchers – could reach different conclusions on the question of detection. Any benefits of mere detection could be offset by the turmoil of the disagreement. A scenario involving more than mere detection could also still create conflict and disorder, but this outcome seems less likely.

**5. Intentional harm to us**

The last scenarios we consider are those in which contact with ETI is harmful to humanity (Figure 1, right column). This is a particularly important set of scenarios because of the strong caution they impose on our SETI and METI endeavors. These scenarios have also received extensive consideration in both fictional and non-fictional realms. Here we explore one main type of scenario in which an ETI could be harmful: intentional harm. The possibility of ETI causing unintentional harm is discussed in the following section. In the intentional harm scenarios, ETI decide that they wish to cause us harm and then follow through on this wish. In the unintentional harm scenarios, ETI do not wish us any harm but inadvertently harm us anyways.

We see two types of scenarios in which ETI might intentionally harm us. The first scenario involves hostile, selfish ETI that attack us so as to maximize their own success. This scenario suggests a standard fight-to-win conflict: a war of the worlds. The second scenario involves ETI that are in no way selfish but instead follow some sort of universalist ethical framework. ETI might attack us not out of selfishness but instead out of a universalist desire to make the galaxy a better place.

## 5.1 Selfish extraterrestrials

A selfish ETI is one that places intrinsic value only on properties of itself: its lives, its welfare, etc. The idea of a selfish ETI is quite prominent in discussions of ETI. For example, geographer Jared Diamond [73], drawing from his expertise in encounters between different intelligent populations on Earth, argues that astronomers are often overly optimistic about ETI encounters:

> The astronomers and others hope that the extraterrestrials, delighted to discover fellow intelligent beings, will sit down for a friendly chat. Perhaps the astronomers are right; that's the best-case scenario. A less pleasant prospect is that the extraterrestrials might behave the way we intelligent beings have behaved whenever we have discovered other previously unknown intelligent beings on earth, like unfamiliar humans or chimpanzees and gorillas. Just as we did to those beings, the extraterrestrials might proceed to kill, infect, dissect, conquer, displace or enslave us, stuff us as specimens for their museums or pickle our skulls and use us for medical research. My own view is that those astronomers now preparing again to beam radio signals out to hoped-for extraterrestrials are naïve, even dangerous.

While Diamond is correct in noting that many astronomers neglect the potential perils of an ETI encounter, it would be a mistake to assume that astronomers are uniformly naïve in this regard. For example, Nobel Laureate astronomer Sir Martin Ryle opposes active efforts to communicate with ETI due to concern that humans would be attacked [36,74-75]. Similar concerns have been raised by several others [26,43,76-77]. Even Carl Sagan, who is usually quite optimistic about ETI encounters, has expressed concern regarding ETI risks [14]. A common theme underlying the pessimism of these various commentators is the likelihood that ETI would be more advanced than humanity.

A core concern is that ETI will learn of our presence and quickly travel to Earth to eat or enslave us. Predation is common among life forms on Earth because it can be more efficient to prey upon other biota than it is to independently utilize autotrophy for energy, carbon fixation, and other nutrients for cellular material [78]. This may be less of a concern if the chirality of organics on Earth is poorly suited as a universal food source [78]. Additionally, an advanced society capable of interstellar travel may be less likely to turn to humans as a source of food or labor because they should have already solved these problems through some combination of machine labor, artificial synthesis, and conservation [14]. Nevertheless, other selfish motives may cause ETI to harm us, such as their drive to spread their beliefs through evangelism (akin to the spread of Christianity or Islam) or their desire to use humans for entertainment purposes. As

Shklovskii and Sagan [14] discuss:

> Or perhaps human beings have some relatively uncommon talent, of which they are themselves entirely unaware. J. B. S. Haldane once pointed out to me that sea lions and seals have a remarkable ability to balance a rubber ball on their noses, which is part of the reason we maintain them in captivity.

Thus in one ETI contact scenario, the ETI use humanity for entertainment purposes just as we use sea lions and seals for this. Shklovskii and Sagan [14] continue to point out that ETI may desire to be the sole galactic power and will eliminate other life forms when they start to get in the way. Similarly, an ETI may simply be interested in using us as a means for growth of their economy. On an individual level they may not be interested in killing us, but may be interested in incorporating us into their civilization so they can sell us their products, keep us as pets, or have us mine raw materials for them. Such a scenario could be harmful or beneficial to us, depending on the methods they use to bring us into their society.

Under what conditions might ETI be self-interested? Here it is again useful to consider possible resolutions to the Fermi paradox, in particular the Sustainability Solution. It is unlikely that humanity will encounter an exponentially expansive civilization [18-19] because we likely would have already detected ETI if exponential expansion could be maintained on galactic scales. Thus exponentially expanding ETI probably do not exist or otherwise do not have the capacity to expand throughout the galaxy. This is fortunate for humanity, since exponentially expansive ETI would likely be quite harmful, just as exponentially expansive populations on Earth (including at least some portions of humanity) can be harmful for other members of their ecosystems. An exception to this is a civilization that has exponentially grown and collapsed in the past but did not succumb to complete ecological collapse. Such a society may recover and choose once again to embark upon a development pattern of exponential expansion. If such an ETI civilization exists today, then they could be extremely harmful, even if they are only moderately more advanced than we are, because if they continue upon their developmental trajectory to rapidly colonize the galaxy, then they will likely consume our resources before their collapse occurs.

As discussed above, we have reason to believe that a sustainable ETI is less likely to be harmful than an unsustainable, exponentially expansive ETI. However, it remains entirely possible for an ETI to be both sustainable and harmful. Such an ETI could be expanding as fast as happens to be sustainably possible, along a colonization wavefront as in the simulations by Newman and Sagan [21]. Unlike the sustainable civilization described above, this *maximally expansive* civilization would be sustainable but still eager to consume whatever resources it could. This type of ETI civilization would likely consume all the resources of Earth and destroy humanity if we got in its way. In the analysis of ETI expansion, a key question is thus whether or not the expansion is occurring at or near the maximal possible rate.

## 5.2 Universalist extraterrestrials

It might seem unlikely that a universalist ETI would intentionally harm us. This is because universalist ETI place inherent value on whatever traits that it values (lives, ecosystems, etc.) regardless of whether it relates to our civilization or theirs. In other words, a universalist ETI civilization would be in no way biased against us. Within humanity, universalism is commonly associated with peace and cooperation, not with harm and destruction. But this is because human populations are all generally similar. If, for example, we seek to maximize total happiness, then we will succeed by avoiding conflict within humanity, because conflict generally reduces happiness for nearly all humans.

Such may not be the case for ETI. Just because an ETI civilization holds universalist ethics does not mean that it would never seek our harm. This is because ETI may be quite different from us and could conclude that harming us would help maximize whatever they value intrinsically [34]. For example, if ETI place intrinsic value on lives, then perhaps they could bring about more lives by destroying us and using our resources more efficiently for other lives. Other forms of intrinsic value may cause universalist ETI to seek our harm or destruction as long as more value is produced without us than with us. Novelist Douglas Adams captures this scenario vividly in *The Hitchhiker's Guide to the Galaxy*, where ETI place intrinsic value on civic infrastructure (or, more likely, on some consequence of its use) and destroy Earth to make way for a hyperspace bypass. At the heart of these scenarios is the possibility that intrinsic value may be more efficiently produced in our absence.

An interesting and important case of universalist ethics in this context is when civilization itself holds intrinsic value. ETI that support this ethical framework would seek to maximize the total number of civilizations, the diversity of civilizations, or some other property of civilizations. All else equal, such ETI would specifically wish for our civilization to remain intact. But all else may not be equal. It is plausible that such ETI might try to harm or even destroy us in order to maximize the number/diversity/etc. of civilizations. This could occur if our resources could be used to more efficiently to generate or retain other civilizations, though this possibility seems highly remote given how efficiently tuned humanity is to its environment. Alternatively, such ETI could seek our harm if they believe that we are a threat to other civilizations.

The thought of humanity being a threat to other civilizations may seem implausible given the likelihood of our technological inferiority relative to other civilizations. However, this inferiority may be a temporary phenomenon. Perhaps ETI observe our rapid and destructive expansion on Earth and become concerned of our civilizational trajectory. In light of the Sustainability Solution to the Fermi paradox, perhaps ETI believe that rapid expansion is threatening on a galactic scale. Rapidly (maximally) expansive civilizations may have a tendency to destroy other civilizations in the process, just as humanity has already destroyed many species on Earth. ETI that place intrinsic value on civilizations may ideally wish that our civilization changes its ways, so we can survive along with all the other civilizations. But if ETI doubt that our course can be changed, then they may seek to preemptively destroy our civilization in order to protect other civilizations from us. A preemptive strike would be particularly likely in the early phases of our expansion because a civilization may become increasingly difficult to destroy as it continues to expand. Humanity may just now be entering the period in which its rapid civilizational expansion could be detected by an ETI because our expansion is changing the composition of Earth's atmosphere (e.g. via greenhouse gas emissions), which therefore changes the spectral signature of Earth. While it is difficult to estimate the likelihood of this scenario, it should at a minimum give us pause as we evaluate our expansive tendencies.

It is worth noting that there is some precedent for harmful universalism within humanity. This precedent is most apparent within universalist ethics that place intrinsic value on ecosystems. Human civilization affects ecosystems so strongly that some ecologists now often refer to this epoch of Earth's history as the anthropocene [79]. If one's goal is to maximize ecosystem flourishing, then perhaps it would be better if humanity did not exist, or at least if it existed in significantly reduced form. Indeed, there are some humans who have advanced precisely this argument [80-82]. If it is possible for at least some humans to advocate harm to their own civilization by drawing upon universalist ethical principles, then it is at a minimum plausible that ETI could advocate harm to humanity following similar principles.

The possibility of harmful contact with ETI suggests that we may use some caution for METI. Given that we have already altered our environment in ways that may viewed as unethical by universalist ETI, it may be prudent to avoid sending any message that shows evidence of our negative environmental impact. The chemical composition of Earth's atmosphere over recent time may be a poor choice for a message because it would show a rapid accumulation of carbon dioxide from human activity. Likewise, any message that indicates of widespread loss of biodiversity or rapid rates of expansion may be dangerous if received by such universalist ETI. On the other hand, advanced ETI may already know about our rapid environmental impact by listening to leaked electromagnetic signals or observing changes in Earth's spectral signature. In this case, it might be prudent for any message we send to avoid denying our environmental impact so as to avoid the ETI catching us in a lie.

## 6. Unintentional harm to us

The harm scenarios considered thus far have all involved ETI that intend to cause us harm, but it is not the only type of scenario in which ETI actually do cause us harm. Specifically, it is possible for ETI to cause us harm despite them not wishing to do so. Here the desires of ETI may even be irrelevant: such ETI could hold any value system from selfish to universalist while still causing unintentional harm in several ways. In one set of scenarios, ETI could inadvertently bring us some sort of physical hazard, such as a disease or an invasive species. In another set of considerations, ETI could inadvertently bring an information hazard, such as technological damage or a presence that demoralizes or destabilizes human society.

**6.1 Physical hazard**

If humanity comes into direct physical contact with either ETI themselves or some ETI artifact, then it may be possible for humanity to be unintentionally harmed. One of the most prominent scenarios of this kind is the transmission of disease to humanity. This scenario is inspired by the many instances in which humans and other species on Earth have suffered severely from diseases introduced from other regions of the planet. Such diseases are spread via the global travels of humans and our cargo and also through certain other disease vectors. Introduced diseases have been extremely potent because the population receiving the disease has no prior exposure to it and thus no build-up of immunity. Indeed, disease introductions are blamed for loss of human life so widespread as to have altered the broadest contours of human history [83].

If ETI could introduce disease to humanity, then the impacts could be – but wouldn't necessarily be – devastating. The disease could quite easily be significantly different from anything our immune systems have ever encountered before. The disease could also be entirely unfamiliar to our medical knowledge, and it could potentially be highly contagious and highly lethal. This combination of contagiousness (i.e. high $R_0$ [84]) and lethality (i.e. high mortality rate) is unlikely in existing pathogens because such pathogens would quickly kill their host population and then die out themselves. Furthermore, if we had already encountered such a disease on Earth, then we likely wouldn't be here anymore. However, a disease from ETI would be new to us. It presumably would not be highly contagious and lethal to the ETI themselves or to the other organisms in their biosphere, but it could be devastating to humans and the Earth system. Then again, ETI biology may be so vastly different from Earth biology that no significant interactions between organisms occur. ETI may have their own contagious diseases that are unable to infect humans or Earth-life because we are not useful hosts for ETI pathogens. After all, the ETI diseases would have evolved separately from Earth biota and thus be incompatible. So while there are reasons to believe that an ETI disease which affected humanity would be devastating, there are also reasons to believe that an ETI disease would not affect humanity.

It is worth noting that a disease brought by an ETI could harm us without infecting us. This would occur if the disease infects other organisms of interest to us. For example, ETI could infect organisms important to our food supply, such as crop plants or livestock animals. A non-human infection would be less likely to destroy humanity and more likely to only harm us by wiping out some potentially significant portion of our food supply. In a more extreme case, ETI disease could cause widespread extinction of multiple species on Earth, even if humans remain uninfected.

It may be possible to protect humanity from diseases brought by ETI. The most straightforward option is simply to prevent contact between the ETI biosphere and Earth's biosphere. Several calls for such prevention have already been advanced, often under the rubric of planetary protection [85]. If we never come into contact with an ETI biosphere, then we probably cannot become infected by its diseases. This fact has implications both for how humanity handles communications with ETI – for example, whether our communications encourage contact–and for human space exploration policy – for example, whether we send probes in search of ETI life, and whether we send these probes back to Earth if life is found.

If prevention fails and ETI disease is contacted, then treatment may be aided by information about the biology of ETI and other organisms in their biosphere. Perhaps such information could be used to develop vaccines or other countermeasures. However, our experience with novel diseases on Earth, such as novel influenza strains, suggests that it takes much less time for a disease to spread than for us to find a cure. The spread of ETI diseases may be even more rapid and the cure even more difficult to develop. Therefore, any head start we can get for our cure development could be highly valuable. This in turn makes remotely received information about ET biology (i.e. biology of the ETI and others in their biosphere) valuable. If we can receive information about ET biology before we make physical contact–for example, if we can receive it via electromagnetic transmission–then perhaps we can develop adequate countermeasures to ET diseases before we encounter them. The possibility that physical contact with ETI may infect humanity with a deadly disease also suggests that we may want to refrain from broadcasting any specifics of our biology. Malicious ETI that learn about our biology will know how to best exploit our immune systems and may even design a human-tailored biological weapon before coming to destroy us. Thus, one possible METI strategy may be to actively seek information about ET biology while carefully guarding the details of human and Earth biology.

Diseases are not the only physical hazard we may unintentionally face from ETI. A similar biological hazard is the invasive species. Whereas a disease infects and harms an organism by overwhelming its immune system, an invasive species affects and harms an ecosystem by overwhelming its ecological functions. The distinction between diseases and invasive species is at most a blurry one. A disease can at least sometimes be classified as an invasive species. Some diseases, such as viral diseases, are not well-classified as species, while some diseases are not invasive because they have a permanent and entrenched status within their host population. Likewise, some invasive species are not diseases per se but instead are harmful in other ways. For example, an introduced predator is a disease only in a metaphorical sense.

In the context of an encounter with ETI, the dynamics of invasive species are similar to the dynamics of introduced diseases. In both cases, humanity is particularly vulnerable due to the extreme novelty of the introduced agent, because our natural defenses and our skilled response efforts are unaccustomed to the agent. Also, in both cases, humanity could benefit from preventing contact with the ET biosphere and from remotely received information about the ET biology. Although an invasive extraterrestrial species seems like it should displace at least some portion of Earth's ecosystem, it is also possible that such invasive species occupy a completely different ecological niche than any extant life on Earth. Thus, we may find that an extraterrestrial invasive species takes up residence on our planet without causing any destruction at all (analogous to a shadow biosphere – see [33,64]).

One non-biological physical hazard that we could face from direct contact with ETI is unintentional mechanical harm. For example, ETI might accidentally crush us while attempting an unrelated maneuver. This scenario parallels instances on Earth in which humans inadvertently destroy the ecosystems of species that then go extinct. All else equal, humanity would generally prefer not causing the extinction of species, but we often prioritize other matters. Indeed, in many cases we may not have even realized that an endangered species was present until after extinction has occurred. Perhaps ETI could inadvertently destroy humanity under analogous circumstances.

In a similar class of scenarios, ETI could inadvertently unleash some harmful force into the galaxy through some act of incompetence, quite possibly harming itself in the process. For example, an otherwise benevolent extraterrestrial civilization could accidentally unleash the extraterrestrial equivalent of an "unFriendly Artificial Intelligence" (uFAI [86]). This ET uFAI would be out of the control of its (benevolent) makers and would likely destroy humanity as it attempted to fulfill whatever objective function it happened to have. The odds that this objective function will happen to benefit humans seems extremely small. Indeed, it may be difficult for humans to create such an objective function even with considerable dedicated effort [86]. In another example, ETI that explore the galaxy using automated self-replicating probes (also known as von Neumann probes) may inadvertently unleash a catastrophic colonization wave that rapidly spreads throughout the galaxy and destroys other civilizations [10,26]. Such a scenario may arise either from faulty design of automated probes or from the malicious intent of artificially intelligent probes. Bostrom [48] suggests that such undesirable outcomes could be the result of evolutionary dynamics in which the undesirables are the strong which survive evolutionary pressures. Finally, it is possible that ETI could render some portion of the galaxy uninhabitable via an accident in a physics experiment, just as there are concerns that certain human physics experiments with particle accelerators could be accidentally destructive [87]. Any of these scenarios would involve the ETI accidentally harming humanity and probably also itself.

**6.2 Information hazard**

If humanity did not come into direct physical contact with ETI, it could still be possible for ETI to unintentionally harm humanity. This could occur if ETI send harmful information to humanity via electromagnetic transmission. A malicious ETI broadcaster could, for example, send a message containing harmful information that either damages human technology, analogous to a computer virus, or coerces humans into a seemingly benign but ultimately destructive course of action, such as the construction of a dangerous device, [76].

As another example, ETI might send information about its biology, perhaps hoping that humanity could use this information to protect itself against ET diseases or invasive species. However, perhaps such an effort would backfire on humanity if we use the information to create a disease, invasive species, or other hazard. The hazard would be created by humans from the information received, and the creation could be intentional or unintentional. But if the creation was intentional, then it would be human intent, not ETI intent. The possibility of an intentional or unintentional informational hazard suggests that at least some care should be taken in efforts to detect and analyze electromagnetic signals sent from ETI.

There is one final information hazard scenario to consider. In this scenario, contact with ETI serves as a demoralizing force to humanity, with strong negative consequences. In human history, contact between modern society and stone age culture usually leads to the demise of the more primitive society. Likewise, in the event of contact with ETI, humanity may be driven toward global cultural collapse when confronted with ETI technology, beliefs, and lifestyle [88]. Even if the ETI are friendly toward us and give us the choice to accept or reject their knowledge, the vast differences between our respective societies may force the more primitive one (ours) into a demoralizing state of societal collapse. For this reason, if ETI do already know of our presence and if they wish to preserve the integrity of our civilization, then they may choose to reveal themselves to us slowly and gradually in order to avoid a calamitous response [23].

**7. Conclusion**

The outcome of contact between humanity and ETI depends on many factors that cannot be fully known at this time. The scenario analysis presented in this paper therefore serves as a means of training our minds to recognize patterns and analyze outcomes before contact with ETI ever occurs. Actual contact may not precisely follow the scenarios considered here, but any amount of analysis to prepare ourselves for contact will increase the likelihood of a positive outcome. Therefore, the analysis presented here serves as a step toward developing a comprehensive strategy for responding to contact with ETI.

Based on the infeasibility of sustained exponential expansion through space, it seems less likely that ETI will destroy us because of their lack of resources. Nevertheless, ETI could still decide to harm us intentionally because of their own ethical considerations, or they may cause us unintentional harm through invasive species or cultural collapse. It is also entirely possible that contact with ETI will have little impact on Earth or humanity, especially if the form of ETI life is vastly different from life on Earth. SETI often assumes that any two intelligent civilizations in the universe could communicate, but we cannot neglect the possibility that the human species will be completely unable to comprehend the language or communication efforts of ETI. The possibility of a neutral ETI encounter, then, is just as worthy of consideration as a scenario with friendly or hostile ETI.

Our analysis suggests some immediate practical recommendations for humanity. One recommendation is that messages to extraterrestrials should be written cautiously. For example, prior messages have included details of human biology, such as the numbers one through ten (our base ten system is likely derived from the number of fingers on our hands) and the form and structure of the DNA molecule. However, details about our biology, though seemingly harmless, may actually help certain ETI to cause us harm. A malicious ETI listener may use a message about human biology to design a potent biological weapon for use against Earth. Since these messages will ultimately be sent toward unknown ETI, we cannot know whether or not they might be received by such a malicious ETI. Therefore, caution is warranted. For example, initial communication with ETI may be best limited to simple mathematical discourse for security purposes until we have a better idea of the type of ETI we are dealing with. In our view, decision making regarding messaging should factor in the probabilities and magnitudes of possible message scenarios through a formal risk analysis that could draw on the scenario analysis presented here.

Another recommendation is that humanity should avoid giving off the appearance of being a rapidly expansive civilization. If an ETI perceives humanity as such, then it may be inclined to attempt a preemptive strike against us so as to prevent us from growing into a threat to the ETI or others in the galaxy. Similarly, ecosystem-valuing universalist ETI may observe humanity's ecological destructive tendencies and wipe humanity out in order to preserve the Earth system as a whole. These scenarios give us reason to limit our growth and reduce our impact on global ecosystems. It would be particularly important for us to limit our emissions of greenhouse gases, since atmospheric composition can be observed from other planets. We acknowledge that the pursuit of emissions reductions and other ecological projects may have much stronger justifications than those that derive from ETI encounter, but that does not render ETI encounter scenarios insignificant or irrelevant.

A final recommendation is that preparations for ETI encounter, whether through METI, SETI, human explorations of space, or any other form, should consider the full breadth of possible encounter scenarios. Indeed, perhaps the central conclusion of the analysis presented here is that ETI contact could proceed in a wide range of ways. It is inappropriate and inadequate to blindly assume that any one specific scenario would result from contact. Until such contact occurs, we simply do not know what would happen. Given the uncertainty, the broad scenario analysis presented here is an important step towards helping us think through and prepare for possible contact.

Despite its merits, our scenario analysis remains fundamentally limited in several important ways. As is common with scenario analysis in general, we offer no quantification scheme for the probabilities of specific scenarios. We also do not quantify the magnitude of the impacts (benefit or harm) of specific scenarios. The result of this is that we are unable to produce a cumulative analysis of the risks and rewards of contact with ETI or attempting to do so with METI. Such a quantitative risk analysis would be of tremendous value for decision making purposes. Indeed, the need has been acknowledged for such analysis in order to inform decisions about METI and other SETI activities [89]. However, the effort required for such an analysis is far beyond the scope of what we can accomplish in a single paper and thus must be left for future work. The scenario analysis presented here is an important step towards a quantitative risk analysis, but it is not a complete risk analysis on its own.

An additional caveat to our scenario analysis derives from the limits of our knowledge about contact with ETI. Because we have no empirical data about ETI, we must extrapolate from the information that we do have available, including knowledge about the observable universe and knowledge about ourselves. We must bear in mind that our observations are inevitably confined to human experience, and so our extrapolations, no matter how generalized, may still contain implicit anthropocentric biases. It is entirely possible that ETI will resemble nothing we have previously experienced or imagined, in which case the contact may not resemble any scenario we could develop. This possibility does not mean that we should completely dismiss any analysis of extraterrestrials, since there is also a strong possibility that the contact would have some resemblance to our scenarios. Nevertheless, the possibility that our experience and imagination could come up severely short reminds us to use caution in interpreting our analysis. Until we actually detect ETI, we will remain highly uncertain as to their nature and to the outcomes that would follow from our contact with them.

One area for future work concerns impacts (benefits and harms) to nonhumans. This paper has focused on the impacts of contact to humanity. We have thus neglected impacts to the ETI, to the rest of Earth, to the rest of the galaxy, and possibly even to other entities as well. We focused on humanity to maintain a reasonably narrow scope for the paper, not because we believe that impacts to nonhumans are unimportant. Indeed, we feel strongly that consideration of impacts to nonhumans represents an important area for future work.

An additional area for future work concerns quantitative risk assessment. A quantitative assessment of the scenarios presented in this paper would be of tremendous use in developing strategies for responding to contact with ETI. However, because we have no observations of ETI, any attempt at quantitative analysis will struggle to assign numerical probabilities to the qualities of an unknown ETI civilization. Certain aspects of this problem, such as rates of expansion and exploration, can be constrained with known physical models, though, so at least some degree of quantification is possible. Additionally, continued exploration of our galaxy and universe will reveal information that will further constrains some of these scenarios such as the distribution of terrestrial planets, the prevalence of Earth-like atmospheric biosignatures, or the existence of artificial radio signals. A complete quantitative assessment of risk from an encounter with extraterrestrials may be difficult to complete in the near future, but even incremental progress will help us choose an optimal strategy if and when we make actual contact with ETI.

Even if contact with extraterrestrials never occurs, our scenario analysis still acts as a set of future trajectories for human civilization. Our thinking about the nature of extraterrestrials and intelligent life in general is really an exercise in imagining the ways that future humans could exist under different circumstances or in different environments. This scenario analysis therefore helps to illuminate the consequences of particular decisions, such as the mode of expansion or the ethical framework of an intelligent civilization, and may help us distinguish between desirable and undesirable trajectories for humanity. As we continue the search for extraterrestrials into the future, perhaps our thinking about the different modes of contact will help human civilization to avoid collapse and achieve long-term survival.

**Acknowledgments**